\documentclass[conference]{IEEEtran}
\IEEEoverridecommandlockouts
\usepackage{cite}
\newcommand{\citet}[1]{\cite{#1}}

\usepackage{amsmath,amssymb,amsfonts}
\usepackage{algorithmic}
\usepackage{graphicx}
\usepackage{textcomp}
\usepackage{xcolor}
\usepackage{multirow}

\def\BibTeX{{\rm B\kern-.05em{\sc i\kern-.025em b}\kern-.08em
    T\kern-.1667em\lower.7ex\hbox{E}\kern-.125emX}}

\usepackage{pifont}% http://ctan.org/pkg/pifont
\newcommand{\cmark}{\ding{51}}%
\newcommand{\xmark}{\ding{55}}%
\def\*#1{\mathbf{#1}}
\def\!#1{\mathbb{#1}}
\def\.#1{\mathcal{#1}}

\definecolor{red}{RGB}{0,0,0}

\begin{document}

\title{Towards Low-Latency Tracking of Multiple Speakers With Short-Context
% Distilled 
Speaker Embeddings
}

% \and
% \and
% \and
% \and
% \and

\author{
\IEEEauthorblockN{Taous Iatariene$^{12}$, Alexandre Guérin$^1$, Romain Serizel$^2$}

\IEEEauthorblockA{
\textit{$^1$
 Orange Innovation}, Rennes, France \\
\textit{$^2$
 University de Lorraine, CNRS, Inria, Loria}, Nancy, France \\
}

taous.iatariene@orange.com, alexandre.guerin@orange.com, romain.serizel@loria.fr
}

\maketitle

\begin{abstract}
% Speaker embeddings are promising identity-related features to complement spatial observations for tracking intermittent speakers, who can move unpredictably while silent. 
Speaker embeddings are promising identity-related features that can enhance the identity assignment performance of a tracking system by leveraging its spatial predictions, i.e, by performing identity reassignment.
Common speaker embedding extractors usually struggle with short temporal contexts and overlapping speech, which imposes long-term identity reassignment to exploit longer temporal contexts.
However, this increases the probability of tracking system errors, which in turn impacts negatively on identity reassignment.
To address this, we propose a Knowledge Distillation (KD) based training approach for short context speaker embedding extraction from two speaker mixtures. We leverage the spatial information of the speaker of interest using beamforming to reduce overlap. 
We study the feasibility of performing identity reassignment over blocks of fixed size, i.e., blockwise identity reassignment, to go towards a low-latency speaker embedding based tracking system.
Results demonstrate that our distilled models are effective at short-context embedding extraction and more robust to overlap. 
Although, blockwise reassignment results indicate that further work is needed to handle simultaneous speech more effectively. 
\end{abstract}


\section{Introduction}
\label{sec:intro}

Speaker tracking refers to the task of estimating the spatial positions of the speakers present in an acoustic scene, given a multichannel audio recording.
It involves performing localization for position estimation~\cite{grumiauxSurveySoundSource2022}, but in the presence of multiple speakers, identity management considerations must be taken into account to assign estimated positions to the correct speakers~\cite{eversLOCATAChallengeAcoustic2020}.
Speaker tracking is often a pre-processing step to other downstream tasks, such as tele-conferencing~\cite{gamperSpeakerTrackingTeleconferencing2012}, automatic speech recognition~\cite{subramanianDeepLearningBased2022}, and speaker separation~\cite{taherianLocationBasedTrainingMultiChannel2022}, which {may impose} a strong low-latency constraint.

Spatial continuity is often assumed in tracking systems~\cite{eversLOCATAChallengeAcoustic2020,quinlanTrackingIntermittentlySpeaking2009,liOnlineLocalizationTracking2019}, which discards the scenario of intermittent speakers, who may move unpredictably during silence. This may lead to the occurrence of speakers with discontinuous spatial tracks, for which many tracking systems fail at maintaining coherent identity assignment~\cite{liOnlineLocalizationTracking2019,iatarieneTrackingIntermittentMoving2025}. Recently, speaker embeddings were proposed to tackle the problem of tracking intermittent and moving speakers, to provide identity-related features for maintaining identity coherence over discontinuous tracks~\cite{iatarieneSpeakerEmbeddingsImprove2025}. 
Given a tracking system, the proposed method leverages on the predicted spatial tracks to perform identity reassignment using speaker embeddings.

Speaker embeddings can be viewed as compact representations capturing information about speaker identity. A common approach for embedding extraction involves models pretrained on large-scale datasets~\cite{desplanquesECAPATDNNEmphasizedChannel2020}, to be used in downstream tasks such as speaker diarization~\cite{dawalatabadECAPATDNNEmbeddingsSpeaker2021}, target speaker extraction~\cite{zmolikova_neural_2023}, and recently for speaker tracking~\cite{iatarieneSpeakerEmbeddingsImprove2025}.
Such pretrained, off-the-shelf models, remain sensitive to speech overlap, notably due to the absence of such scenarios in their training datasets~\cite{jungSearchStrongEmbedding2023}. 
They usually involve temporal pooling of frame level latent features~\cite{desplanquesECAPATDNNEmphasizedChannel2020,snyderXVectorsRobustDNN2018}, and often require long temporal context to reach invariance to phonetic and linguistic content~\cite{jakubecDeepSpeakerEmbeddings2024}.
% from very short inputs challenging~\cite{zeinali_short-duration_2021}.
% This makes pretrained speaker embedding extractors ill-suited for speaker tracking, which require features derived from short and noisy audio inputs, often containing overlapping speakers.

{In consequence}, Iatariene \textit{et al.}~\citet{iatarieneSpeakerEmbeddingsImprove2025} considers performing identity reassignment at the fragment-level, where a fragment is defined as a temporal segmentation 
of spatial tracks into periods of activity, thus allowing long temporal context for speaker embedding extraction.
% Not only does this increase the overall system's latency, but it also assumes long term reliability from the considered tracking system, where spatial identity coherence needs to be guaranteed at the fragment level, i.e, that a fragment can be linked to only one speaker. 
{Not only does this increase the overall system's latency, but it also leads to potentially more speech overlap. This impacts negatively on embedding extraction, despite the use of beamforming to mitigate overlap~\cite{iatarieneSpeakerEmbeddingsImprove2025}.
Furthermore, the method assumes long term reliability from the considered tracking system, where spatial identity coherence needs to be guaranteed at the fragment level, i.e, that a fragment can be linked to only one speaker.}
This strong assumption may not always be verified. The longer the temporal view, the higher the probability of tracking errors such as swaps or breaks~\cite{eversLOCATAChallengeAcoustic2020}. 
% Moreover, longer temporal contexts for embedding extraction lead to the increased presence of speech overlap, which impacts negatively embedding extraction~\cite{iatarieneSpeakerEmbeddingsImprove2025}.
% Despite the use of beamforming prior to embedding extraction, larger drop in performances when given longer temporal context for embedding extraction compared to ideal separation was noticed sub-optimal performances were noticed on the reassignment system on 2 speaker scenes, with increased decay on longer input durations

% This increased the systems latency but 
% Beamforming was applied before embedding extraction by leveraging on the multichannel audio and spatial information provided by the considered tracking system.
% and increased the system's latency by considering fragment levelwhile considering long- . However, the result of beamforming separation remains imperfect, since it depends notably the accuracy of the spatial information provided by the tracking system. This has a negative impact on speaker embedding extraction.
% Also, the long temporal context needed for embedding extraction requires longer and coherent spatial information, which imposed to perform identity reassignment on long-term 
% comes at the cost of increased latency, and higher probability of spatial errors by the tracking system.

{This paper aims at going towards a low-latency speaker embedding based tracking system.
First, inspired by the works on other speaker embedding extractors for short duration speaker verification~\cite{sangOpenSetShortUtterance2020} and
% speaker recognition~\cite{pengLabelfreeKnowledgeDistillation2022}, 
multi-speaker embedding extraction~\cite{cord-landwehrTeacherStudentApproachExtracting2023}, we propose a Knowledge Distillation (KD) approach for short context speaker embedding extraction, with increased robustness to speech overlap. In KD based training, a \textit{student} model is trained to reproduce the robust latent space of a \textit{teacher} model pretrained on a large-scale dataset. {The training framework involves providing the student model with short temporal \textit{crops} of various durations, and leverages beamforming for increased robustness against overlapping speech.}
Second, we introduce blockwise identity reassignment, which processes over temporal blocks of fixed size. This reduces the system's latency to the chosen block duration, and removes the assumption of spatial identity coherence on fragment level.}
% We propose an identity reassignment system that does not rely on the notion of fragments, thereby eliminating the need for spatial identity coherence at the fragment level. Instead, w
% Second, we propose a training framework for short context speaker embedding extraction, with increased robustness against speech overlap.
% o reduce the constraint of long term spatial reliability from the tracking system and alleviate the sensitivity to speech overlap. 

% where the pretrained \textit{teacher} model provides robust speaker embeddings as targets for training a \textit{student} model which tries to reproduce the targets given very short and noisy inputs. 
% We make it context-aware by proving it with beamformed inputs during training, and
% \textcolor{red}{We increase the model robustness to speech overlap by using training data containing overlap.}
% and experiment with VAD information for even more guided embedding extraction~\cite{horiguchiGuidedSpeakerEmbedding2025}

We evaluate the efficiency of our student speaker embedding extractor, against the pre-trained teacher model, for speaker tracking. We first use fragment-level identity reassignment~\cite{iatarieneSpeakerEmbeddingsImprove2025} to assess the interest of KD in our approach.  
We investigate the feasibility of blockwise identity reassignment by analyzing its sensitivity to speech overlap across various block sizes, and compare its performance to the fragment-level approach.

\section{Related works}
\label{sec:related}

\subsection{Speaker tracking}
\label{sec:related_tracking}

We consider a multichannel audio recording of an indoor acoustic scene (e.g, office), containing a number of speakers who may move around while speaking or when silent. 
Speaker tracking aims at reconstructing the time-varying position of each individual speaker present in the recording, which we refer to as \textit{tracks}. Each track can be described by its track identity, and its set of time-varying positions. 
% Ideally, the number of tracks should be equal to the number of speakers in the scenes, although 

The standard approach for tracking has long been to follow a two-step process of source localization followed by Bayesian filtering (e.g, Kalman filter), and to perform observation-to-track association in the multi-speaker case~\cite{eversLOCATAChallengeAcoustic2020}. 
Most of the developed methods consider continuous movement during inactivity periods~\cite{liOnlineLocalizationTracking2019,quinlanTrackingIntermittentlySpeaking2009}, setting aside the case in which speakers move unpredictably while silent. The spatial track discontinuity induced by such scenario makes it difficult to maintain a coherent track identity over time~\cite{liOnlineLocalizationTracking2019,iatarieneTrackingIntermittentMoving2025}. 

Recently, deep learning based tracking approaches have framed tracking as a task of ordered localization,
relying on Permutation Invariant Training (PIT) to reach this objective~\cite{diaz-guerraPositionTrackingVarying2023a,subramanianDeepLearningBased2022}.
However, a known limit to PIT-trained models is the \textit{block permutation problem}~\cite{chenContinuousSpeechSeparation2020}. 
Calling a PIT-trained model multiple times on successive audio blocks (e.g for online processing) can cause order confusion at the inter-block level, i.e, the chosen order of predictions may vary across blocks.
This makes predicted tracks coherent only at the block level. Otherwise, arbitrary permutations between speakers may affect the spatial identity coherence.

\subsection{Speaker embeddings}
\label{sec:related_emb}

Speaker embeddings are low-dimensional features that encapsulate information about speaker identity.
They are usually obtained by training a neural network over a large-scale dataset containing a great variety of speaker identities, using supervised learning through classification~\cite{snyderXVectorsRobustDNN2018,desplanquesECAPATDNNEmphasizedChannel2020}, or unsupervised learning~\cite{stafylakisSelfsupervisedSpeakerEmbeddings2019a}.
This enables the learning of a latent space in which embeddings from the same speaker are close, while embeddings from different speakers are well separated.
Neural network architectures, such as x-vectors~\cite{snyderXVectorsRobustDNN2018}, usually comprise frame-level feature extractors, followed by temporal pooling for robust embedding extraction~\cite{jakubecDeepSpeakerEmbeddings2024}.

Such models perform best when the speaker embedding extraction closely matches the conditions of the training setup. 
Usually designed for the task of speaker verification~\cite{varianiDeepNeuralNetworks2014,snyderXVectorsRobustDNN2018,desplanquesECAPATDNNEmphasizedChannel2020}, these models require long temporal contexts of several seconds to be invariant to phonetic and linguistic content~\cite{jakubecDeepSpeakerEmbeddings2024}. They also remain sensitive to the presence of speech overlap, making more challenging the direct use of such models for applications containing overlapping speakers, such as diarization~\cite{dawalatabadECAPATDNNEmbeddingsSpeaker2021,jungSearchStrongEmbedding2023}.

\section{Short-context speaker embeddings for tracking}
\label{sec:training_kd}

Speaker embeddings have only recently been employed for speaker tracking, addressing the problem of spatial track discontinuity when having intermittent and moving speakers~\cite{iatarieneSpeakerEmbeddingsImprove2025}. Identity-related observations become important in this context, due to the occurring of discontinuous spatial tracks. 

\begin{figure}
    \centering
    \includegraphics[width=0.95\linewidth]{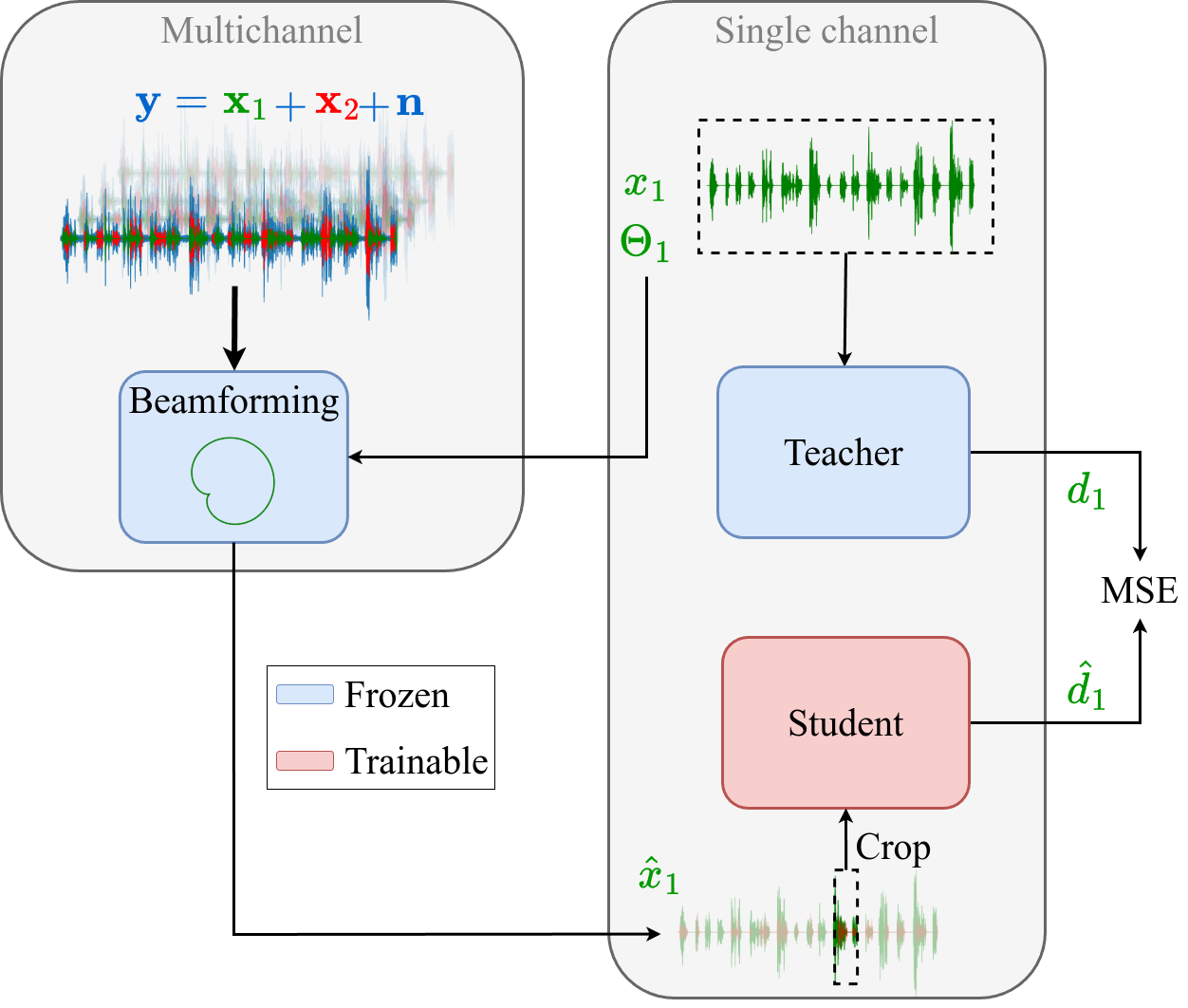}
    \caption{Overview of the Knowledge Distillation (KD) based training approach for short context speaker embedding extraction from a two speaker mixture.}
    \label{fig:schema}
\end{figure}

We propose to train a speaker embedding extractor tailored for the downstream task of speaker tracking. It can therefore leverage on estimated spatial tracks that can be provided by any tracking system. 
We focus on the two following features: short temporal context extraction, and robustness to speech overlap. 
% To increase the robustness of embedding extraction for the downstream task of speaker tracking, we propose to train a speaker embedding extractor with two important features: short temporal context extraction, and robustness to speech overlap. 
To obtain a model able to extract the speaker embedding of one of the speakers composing a mixture, given short temporal context and the prior knowledge of its spatial track, we propose a Knowledge Distillation (KD) based training approach, with an overview given in Fig.~\ref{fig:schema}. 
% We propose to train a speaker embedding extractor targeted for the downstream task of speaker tracking.  
% It can therefore leverage on estimated spatial tracks that can be provided by any tracking system. 
% We focus on two important features needed for a speaker embedding model in such application: short temporal context extraction, and robustness to speech overlap. 

% The context-aware speaker embedding extractor aims at estimating the speaker embedding of one of the two speakers composing the mixture, given prior information about it, and short temporal context. In our case, prior information about the speaker is its time-varying spatial track, which can be estimated by a tracking system.

% If the target speaker is speaker 1, and $\Theta_1$ denotes its corresponding track, then the speaker embedding extraction model $\*M$ estimates a speaker embedding $\*{\hat{e}}_1$, given only several frames of speaker activity. It exploiting  the prior knowneldge of $\Theta_1$. i.e, $\*{M}(\*y, \mathcal{P}_1) = \*{\hat{e}}_1$, given only several frames of speaker activity.

\subsection{Distillation based training}
\label{sec:training_kd_sub}

In KD based training, instead of directly optimizing the training towards a speaker classification task, as done usually for supervised embedding extractors training~\cite{jakubecDeepSpeakerEmbeddings2024}, we leverage on the latent space learned by a pretrained \textit{teacher} model to provide targets for training a \textit{student} model. Such latent space already verifies the properties of small intra-speaker and large inter-speaker distances between embeddings, and simplifies the learning objective of short context and overlap-robust embedding extraction.
As such, the student speaker embeddings should ideally remain invariant to phonetic and linguistic content and be robust to the presence of noise and residual signal from a concurrent speaker, despite the short temporal context taken as input.
% Initially used for model reduction purposes~\cite{pengLabelfreeKnowledgeDistillation2022}, this technique has been recently applied for multi-speaker~\cite{cord-landwehrTeacherStudentApproachExtracting2023} and frame-wise embedding extraction for speaker diarization~\cite{cord-landwehrFrameWiseOverlapRobustSpeaker2023}.

We consider a multichannel mixture signal $\*y =  \*{x}_1 + \*{x}_2 +  \*{n}$ of two speakers, where $\*x_1$ and $\*x_2$ are the reverberated (\textit{wet}) speeches of the two speakers composing the mixture, $\*n$ the noise. The speakers are described by their spatial track, $\Theta_1$ and $\Theta_2$.
One mixture $\*y$ can provide two items of training data, depending on the target speaker. We suppose in the following that the target speaker is speaker~1. We note as $\*T(\cdot)$, $\*S(\cdot)$, the teacher and student models, respectively.

\subsubsection{Teacher}
\label{sec:training_kd_sub_teacher}

The role of the teacher is to provide a training target embedding for the student model, noted $d_1$. 
The teacher is pretrained over a large-scale dataset with a wide range of speaker identities, to learn a robust speaker embedding latent space. Its weights are kept frozen during the KD training.
The target embedding $d_1$ is obtained by leveraging on the multichannel wet speech signal $\*x_1$ prior to mixture computation. One channel from $\*x_1$ is chosen for embedding extraction, noted $x_1$, therefore $d_1 = \*T(x_1)$.

\subsubsection{Student}
\label{sec:training_kd_sub_student}
To predict an estimation ${\hat{d}}_1$ given the mixture $\*y$, the student speaker embedding extractor relies on a beamforming algorithm $\mathcal{B}$, which is a spatial filtering technique. It leverages spatial information $\Theta_1$ to combine the multiple channels of the mixture $\*y$ into an enhanced output ${\hat{x}}_1 = \mathcal{B}(\*y, \Theta_1)$.
The student speaker embedding extractor provides an estimated embedding ${\hat{d}}_1 = \*S({\hat{x}}_1)$, which ideally has an increased robustness to speech overlap thanks to beamforming.

% \subsubsection{Training objective}
% Similar to~\citet{cord-landwehrTeacherStudentApproachExtracting2023}, a MSE loss function is computed between the student's embedding prediction ${\hat{d}}_1$ and the teacher's target $d_1$. 

\subsection{Temporal context}
\label{sec:training_kd_tmp}

To reach the goal of embedding extraction with short temporal context, short durations need to be given to the student model during training.
A random crop along the beamformed signal ${\hat{x}}_1$ is made prior to embedding estimation. 
We vary both the crop duration and crop start for each training item to ensure increased robustness and generalization to diverse temporal contexts.
It is made sure that all crops contain enough activity from the target speaker, by monitoring the voice activity along the crop.

The teacher model, on the opposite, uses the whole signal duration of $\*x_1$. 
This ensures optimal targets for KD based training, thanks to the use of long temporal context.
The difference in duration between the teacher and student model is precisely what enables the student to learn to produce embeddings that are as robust as those of the teacher, but using shorter temporal contexts.

% \subsubsection{Contrastive loss}

% \textcolor{red}{[TODO: Another loss ?]}

\section{Speaker embeddings for identity reassignment}
\label{sec:identity_reassignment}

We consider the predicted tracks of any tracking system, described by their track identities and temporal positions. 
Identity reassignment enables the update and correction of the attributed identities, to reduce associations errors. 
We describe in Sec.~\ref{sec:identity_reassignment_frag} and Sec.~\ref{sec:identity_reassignment_block} fragment-level~\cite{iatarieneSpeakerEmbeddingsImprove2025} and the proposed blockwise speaker embedding based identity reassignment systems, respectively.

\subsection{Fragment-level identity reassignment}
\label{sec:identity_reassignment_frag}

Recently, to cope with the need of long temporal pooling for embedding extraction, identity reassignment at a so-called {fragment} level was proposed~\cite{iatarieneSpeakerEmbeddingsImprove2025}.
A fragment is defined as a temporal segmentation of spatial tracks into periods of activity, which usually ensures several seconds of temporal context for embedding estimation. 
% Beamforming, a spatial filtering technique that can leverage on spatial track information, is applied prior to fragment embedding extraction to reduce sensitivity to speech overlap.
Identity reassignment processes fragments by temporal order of appearance.
A fragment embedding is estimated through a pretrained embedding extractor~\cite{desplanquesECAPATDNNEmphasizedChannel2020}.
It is then compared to a pool of enrollment embeddings. These are precomputed before reassignment to ensure robustness and to represent the available speaker identities effectively. 
The identity of the enrollment embedding with the highest cosine similarity to the fragment embedding is assigned as the fragment’s new identity.

% This also has a negative impact on beamforming, which leverages on the fragment spatial information for enhancement.

\subsection{Blockwise identity reassignment}
\label{sec:identity_reassignment_block}

Performing identity reassignment at the fragment level has a primary noticeable limitation: it increases the latency of the system. Moreover, it assumes spatial identity coherence at the fragment level, that is, a fragment can be attributed to one speaker at most. This assumption doesn't always hold, especially with a PIT-trained neural tracker, which can suffer from the block permutation problem. 

To overcome such issues, we propose to perform the identity reassignment method described in Sec.~\ref{sec:identity_reassignment_frag} over blocks of fixed duration, which we refer to as 
blockwise identity reassignment. 
Instead of segmenting spatial tracks into fragments, this approach processes temporal blocks, sequentially in a streaming fashion, regardless of the signal content, including variations in speaker activity.
Each block contains portions of estimated tracks. As previously, they are processed by temporal order of appearance within the block.
% The same identity reassignment method as described in Sec.~\ref{} is followed.
% Within each block, portions of tracks are processed by temporal order of appearance, as previously. 
One speaker embedding per portion of track is extracted,
% obtained using beamforming and a speaker embedding extractor. 
which is then compared to a pool of enrollment embeddings to take a reassignment decision.

Blockwise identity reassignment allows going towards a low-latency tracking system using speaker embeddings, where the latency is defined as the block size used for reassignment. 
It also relaxes the assumption of spatial identity coherence at the fragment level, \textcolor{red}{although maintaining this assumption over shorter durations at the block level}. 
Consideration must be given to the choice of the block size. Longer block sizes increase the risk of permutation errors, especially if spatial discontinuities occur within a block. Shorter block sizes reduce this risk, since the speakers are less likely to move significantly on small durations. They also reduce the system's latency. However, they result in shorter temporal contexts for embedding extraction, which is already known to degrade embedding quality, and impacts negatively identity reassignment~\cite{iatarieneSpeakerEmbeddingsImprove2025}.

\section{Experimental design}
\label{sec:exp_design}

We evaluate the proposed speaker embedding extractor for its downstream application: speaker tracking, using both fragment level and blockwise identity reassignment.
% The backbone tracking system over which identity reassignment is being performed is a neural tracker described in Sec.~\ref{sec:exp_neuraltracker}. 
% For evaluation, we rely on speaker embedding based identity reassignment at the fragment level~\cite{iatarieneSpeakerEmbeddingsImprove2025}, detailed in Sec.~\ref{sec:related_frag}. 
% To go towards a low-latency speaker embedding based tracking system, we study the impact of relaxing the assumption of spatial identity coherence at the fragment level by performing blockwise identity reassignment.

\subsection{Datasets and metrics}
\label{sec:dataset_metrics}

The training dataset for both neural tracker and speaker embedding extraction consist in synthetic two speaker scenes in the First Order Ambisonics (FOA) multichannel audio format~\cite{iatarieneSpeakerEmbeddingsImprove2025, iatarieneTrackingIntermittentMoving2025}.
Items of training data are generated on-the-fly during training. 
Speech utterances are taken from the LibriSpeech train-clean-100 and train-clean-360~\cite{panayotovLibrispeechASRCorpus2015}, forming a pool of 1172 different speaker identities.

The LibriJump dataset~\cite{iatarieneTrackingIntermittentMoving2025} is used for evaluation. 
It consists in three evaluation subsets of 150 acoustics scenes in the FOA format, of 60~s each, with a SNR of 15~dB. There is one subset per maximum number of speakers (1, 2 or 3) in the scenes. It contains intermittent speakers who change position during silences, thus having discontinuous spatial tracks. 
We conduct experiments on the LibriJump 1spk and 2spk subsets, \textcolor{red}{which have a maximum number of speakers per scene of 1 and 2, respectively}.
{The 1spk subset allows focusing on minimal temporal context embedding extraction without the presence of speech overlap. The more challenging 2spk subset allows for further evaluation of the robustness against speech overlap.}

We use the tracking association accuracy (AssA) score, which provides an overall score to evaluate identity assignment~\cite{luitenHOTAHigherOrder2021,iatarieneTrackingIntermittentMoving2025}. 
% AssA provides an overall score to evaluate identity assignment. 
% This is to put in contrast to the previously popular identity switches (IDSW) metric, which compares ground truths and predictions at the frame level. 
{We present the mean AssA scores obtained after running a 80-20\% bootstrap on the evaluation datasets. Standard deviation did not exceed 1\% in all experiments.}

\subsection{Speaker embedding extractor}
\label{sec:exp_training_ts}

We rely on state-of-the-art ECAPA-TDNN architecture for both of teacher and student models~\cite{desplanquesECAPATDNNEmphasizedChannel2020}. It leverages Time-Delay-Neural-Networks and Squeeze-and-Excitation modules for frame-level feature extraction, and performs attentive statistics pooling to estimate an embedding of dimension~192.

The teacher model is the Speechbrain~\cite{ravanelliSpeechBrainGeneralPurposeSpeech2021} pre-trained version of this architecture\footnote{https://huggingface.co/speechbrain/spkrec-ecapa-voxceleb}, described in particular by its number of channels which is 1024. It is trained on the large-scale datasets Voxceleb1 and Voxceleb2~\cite{nagraniVoxCelebLargeScaleSpeaker2017} containing 7205 speaker identities using the Additive-Angular-Margin (AAM) softmax loss function.
The student architecture mirrors the teacher's. We do experiments in which we reduce the number of channels from 1024 to 512. We train the student model from scratch or from the teacher weights, to see if it can benefit from non-random initial weights.

Similar to Cord-Landwehr \textit{et al.}~\citet{cord-landwehrTeacherStudentApproachExtracting2023}, we choose the MSE loss function as a training objective between the student's embedding prediction ${\hat{d}}_1$ and the teacher's target $d_1$. 
To assess the benefit of KD based training for short context speaker embedding extraction, we run experiments training the student models with a classification objective using the AAM-softmax loss (similar to the teacher's pretraining objective) but on the same training dataset containing beamformed crops of various durations with potentially overlapping speakers.

As for the processing of inputs during training, described in Sec.~\ref{sec:training_kd_sub_teacher}, the omnidirectionnal channel $W$ of the FOA audio format is given for the teacher target embedding computation.
Beamforming and random cropping of the student's inputs are performed prior to the student's embedding extraction. We use the Delay-and-Sum beamformer adapted to the FOA audio format~\cite{baqueAnalyseSceneSonore2017}.
During training, random crop durations are set to 250, 500, 750, 1000, 1500, 2000, 8000~ms, as we empirically found that models trained only on very short inputs durations were either under-performing or not converging.

\subsection{Tracking system}
\label{sec:exp_neuraltracker}

The neural tracker is based on the source splitting approach~\cite{subramanianDeepLearningBased2022}. 
 The neural network architecture consists in a CRNN with a projected BiLSTM, which predicts splitting masks in the latent multi-speaker space, to obtain ordered position estimates on a discrete spherical grid using a PIT training approach. 
 The maximum number of simultaneous predictions $N$ is {fixed} by the number of output branches after splitting (two in our experiments).

\subsection{Identity reassignment}
\label{sec:exp_reassignment}

Identity reassignment relies on enrollment embeddings, as well as fragment or blockwise embeddings, depending on the method.
We use two enrollment embeddings, for both fragment-level and blockwise identity reassignment methods, and for all the experiments, which is equal to the number of speakers present in the scenes.
Enrollments are extracted following the same procedure as Iatariene \textit{et al.}~\citet{iatarieneSpeakerEmbeddingsImprove2025}.

To remain close to the training framework, the FOA Delay-and-Sum beamformer~\cite{baqueAnalyseSceneSonore2017} is used to extract speaker embeddings for reassignment.
For the blockwise method, we vary the block sizes and therefore the temporal context for embedding extraction from short sizes (8, 25, 50 frames), to longer ones (100, 200 frames) in which the tracking system may be more prone to permutation errors. 
\textcolor{red}{Given the frame size of 32~ms, it corresponds to 256, 800, 1600~ms and 3200, 6400~ms.}
Fragment embeddings are extracted given temporal contexts of 250, 750, 1500~ms, similar to Iatariene \textit{et al.}~\citet{iatarieneSpeakerEmbeddingsImprove2025}. We denote as \textit{whole} the case where the entire fragments are used.

\section{Results}
\label{sec:results}

We start by studying the distilled speaker embedding extractor, using the fragment level identity reassignment system {in Sec.~\ref{sec:res_short}}.
We move to blockwise identity reassignment in Sec.~\ref{sec:res_blockwise}. We analyze the influence of block size on performance, and compare it to fragment-level reassignment. 

\subsection{Short-context speaker embeddings}
\label{sec:res_short}

Tab.~\ref{tab:tab1_ts} displays the AssA scores obtained after performing fragment-level reassignment over the neural tracker's predictions, on the LibriJump 2spk subset. Several temporal contexts, starting from the fragment beginnings, and several embedding extractors are compared. 
As a reference, the baseline performance before reassignment is of 38.1~\%.

\begingroup
\setlength{\tabcolsep}{5pt} % Default value: 6pt
\renewcommand{\arraystretch}{1} % Default value: 1
% \begin{table}[!htbp]
\begin{table}[]
\centering
\centering
\caption{{
% Mean AssA scores for several speaker embedding extraction models, on $J=2$ speaker scenes, using the fragment level identity reassignment method~\cite{iatarieneSpeakerEmbeddingsImprove2025}. Number of enrollments $M=J=2$
% Notations: T = Teacher, S = Student, FS = from scratch, C = number of channels, KD = knowledge distillation training, AAM = Additive Angular Margin classification based training.
Fragment-level identity reassignment scores on the LibriJump 2spk dataset. C = channel number. FS = trained from scratch. KD = knowledge distillation. Baseline = before reassignment.
}}

\label{tab:tab1_ts}
\begin{tabular}{ccccccccc}

 % \cline{1-7} \cline{9-9}
 \hline
\multirow{2}{*}{Model} & \multirow{2}{*}{C} & \multirow{2}{*}{FS} & {\multirow{2}{*}{KD}} & \multicolumn{3}{c}{Temporal context {[}ms{]}} & \multirow{7}{*}{\vrule height 16ex width 0.4pt} & \multirow{2}{*}{Baseline} \\ \cline{5-7} 
  &  &  & \multicolumn{1}{c}{} & 250 & 750 & \textit{whole}   &   & \\ \hline \hline 
% \multicolumn{4}{c}{Before reassignment} & \multicolumn{3}{c}{38.1\%} \\
% \hline
Teacher & {1024} & {$-$} & {$-$} & 46.8 & {61.3} & {63.9} & & \multirow{5}{*}{38.1}\\ \cline{1-7} 
\multirow{4}{*}{Student} & {1024} & \multirow{2}{*}{\cmark} & \multirow{2}{*}{\cmark} & {53.1} & 58.3 & 61.7  & & \\
 & {512} & &  & 51.9 & 57.7 & 63.4  & & \\ \cline{2-7} 
 & \multirow{2}{*}{1024} & \multirow{2}{*}{\xmark} & \cmark & \textbf{54.8} & \textbf{63.1} & \textbf{65.5}  &&  \\
 &  &  & \xmark & 38.9 & 38.2 & 39.7  && \\  
 \hline
 % \cline{1-7} \cline{9-9}
\end{tabular}
\end{table}
\endgroup

{
After reassignment, AssA increases to 63.9~\% using the teacher model and \textit{whole} fragment durations. AssA reaches the highest value of 65.5~\% when using the student model initialized from the teacher's weight with 1024 channels. 
On shorter durations, the difference is larger: using the shortest 250~ms temporal context, AssA achieves 46.8~\% and 54.8~\% using the teacher and student models, respectively, which corresponds to an 8~\% absolute increasing.}

On the following, we study the impact of several factors on the short context embedding extractor training, namely : the use of distillation, weight initialization, and model size. 

\subsubsection{Interest of distillation}
% Since all AAM trained models underperform the KD ones, only one of them was displayed in Tab.~\ref{tab:tab1_ts}, using the student trained from the teacher's weights with 1024 channels. 
Comparing the model trained with AAM-softmax to its corresponding KD one (model trained from the teacher's weights with 1024 channels, i.e., two last rows of table~\ref{tab:tab1_ts}), we notice that AssA goes from an absolute 15.9\% drop on the shortest 250~ms temporal context, to a 25.8\% drop with \textit{whole} temporal context.
{This shows the interest of KD for our application, in that it simplifies the task of learning a robust latent space from minimal context}.
% In the following, we won't consider any longer the AAM models.

\subsubsection{Weight initialization}

If all the student models perform better than the teacher when using short temporal context of 250~ms, the teacher's performance out passes the students most of the time when using longer contexts.
% , except for the student trained from the teacher's weights with 1024 channels, which outpasses the teachers.
\textcolor{red}{We notice that models trained from scratch mainly learn to optimize performance on short temporal segments and underperform the teacher on longer contexts.
The model initialized from the teacher’s weights starts from a configuration suited for long temporal embedding extraction. It clearly retains the ability to process longer contexts effectively, as it achieves the best reassignment scores even when given long temporal inputs.
}
% Ca ne parait pas très surprenant :
% * from scratch apprend essentiellement sur des courts
% * From teacher weight part d'un modèle qui fonctionne bien sur des temps longs et le règle pour fonctionner sur des temps courts (en essayant de ne pas "oublier" comment fonctionner sur des temps longs).

% This leads to conclude that non-random weights initialization is beneficial to the student's training.
% on longer temporal contexts leads to conclude that distillation from the teacher's weights helps in keeping the best of both worlds, i.e. the ability to exploit long contexts, and increased behavior on shorter ones.

\subsubsection{Model size}
The last observation derived from the experiment with a smaller channel size (1024 versus 512), with random weight initialization, is more of an opening and would deserve further analysis. Nevertheless, it indicates that reducing the model size from 1024 to 512 does not have a significative impact on the performance, with the highest absolute difference being of 1.7~\% using \textit{whole} temporal contexts.
{This result calls for the design of a more compact architecture that could go without a significative loss of performance for the task of identity reassignment for speaker tracking.}
% The result is somewhat logic, as our task of reassignment with few identities is less complex than identifying a speaker among thousands, hence not necessitating such a high-dimensional space.

\subsection{Blockwise identity reassignment}
\label{sec:res_blockwise}

{In this section, we propose to conduct experiments using blockwise instead of fragment-level identity reassignment. We retain for next the best performing student model, initialized from the teacher weights with 1024 channels.}
Tab.~\ref{tab:tab2_blockwise} displays the AssA scores for the blockwise identity reassignment method on LibriJump 1spk and 2spk subsets, using both student and teacher models. Several block sizes, going from 256~ms to 6.4~s are used. Scores prior to reassignment were of 55.4~\% and 38.1~\% on the 1spk and 2spk subset, respectively.

\subsubsection{Sensitivity to speaker overlap}
\label{sec:res_block_overlap}

{A first observation can be made about the significant drop in performance when dealing with two speakers as opposed to one speaker, even prior to reassignment. 
This highlights the high sensitivity of both models to speech overlap. On one hand, this degrades the quality of embedding extraction. On the other hand, this degrades the quality of spatial track estimation, which in turn impacts on beamforming and thus on embedding extraction.}
% \textcolor{red}{A similar amount of decay of approximately 30\% can be observed for both models, which indicates that the student isn't any better than the teacher to deal with speaker overlap.}

\subsubsection{Sensitivity to block permutation problem}
It is interesting to note that, on the 2spk subset, both models exhibit bell-shape scores. After increasing, AssA decreases starting from 800~ms and 1600~ms block size for the student and teacher, respectively. 
This is not observable on the 1spk subset, whereas despite almost a 50-50 split of predictions over the two output branches of the tracker (as showed by an AssA score of 55.4~\% prior to reassignment), the identity reassignment performance keeps increasing.
The bell-shaped behavior on the 2spk subset therefore gives a clear indication about the duration from which the neural tracker is more likely to be prone to permute the speaker's tracks within a block over its two output branches.
\textcolor{red}{It further justifies the need for short context embedding extraction, to avoid relying on long temporal inputs, over which the tracking system may fail.}

\subsubsection{Student versus teacher performance}

Comparing the student and teacher models, the same conclusion may be observed as from fragment-level experiment. Indeed, as expected and for both datasets, the student performs better on the shortest 250~ms context, while the teacher is better on the longest temporal contexts of 6400~ms. On the mid-length temporal contexts of 800, 1600~ms and 3200~ms, using the 2spk subset, we observe that the student performs better than the teacher model. This suggests that, although both models are sensitive to speech overlap (as noted in Sec.~\ref{sec:res_block_overlap}), the student model demonstrates greater robustness in managing such overlap.

% Surprisingly, the trend is reversing with longer contexts and teacher performs significantly better, even on the 1spk subset where there is no presence of speech overlap. Scores go up to a 13.4\% difference on the 3200~ms block size on the 2spk subset. 
% This contrasts with previous results on fragment-level reassignment where performance on longer durations where rather similar. 
% \textcolor{red}{This can be explained by the higher number of reassignment decisions that need to be taken when performing blockwise reassignment, coupled with the general increased robustness of the teacher to extract more robust speaker embedding when given longer durations.}
\begingroup
\setlength{\tabcolsep}{7pt} % Default value: 6pt
\renewcommand{\arraystretch}{1} % Default value: 1
\begin{table}[]
\centering
\caption{Blockwise identity reassignment performance on 1spk and 2spk subsets for teacher and student models.}
\label{tab:tab2_blockwise}
\begin{tabular}{ccccccc}
\hline
\multirow{2}{*}{Model} & LibriJump &  \multicolumn{5}{c}{Temporal context {[}ms{]}} \\ \cline{3-7} 
{} & subset  & 256 & 800 & 1600 & 3200 & 6400 \\ \hline \hline
\multirow{2}{*}{Student} & 1spk  & \textbf{82.1} & 91.0 & 94.0 & 96.3 & 97.6 \\
 & 2spk  & \textbf{49.4} & \textbf{61.6} & \textbf{61.2} & \textbf{56.9} & 49.9 \\ \hline 
\multirow{2}{*}{Teacher} & 1spk & 72.3 & \textbf{92.8} & \textbf{96.2} & \textbf{98.4} & \textbf{99.2}\\
 & 2spk  & 40.0 & {54.6} & {56.1} & {55.0} & \textbf{50.0}
\end{tabular}
\end{table}
\endgroup

\begingroup
\setlength{\tabcolsep}{10pt} % Default value: 6pt
\renewcommand{\arraystretch}{1} % Default value: 1
\begin{table}[!b]
\centering
\caption{Comparison of blockwise and fragment-level identity reassignment with the student model on LibriJump 2spk subset.}
\label{tab:tab3_start}
\begin{tabular}{ccccc}
\hline
\multirow{2}{*}{{Reassignment}}  & \multirow{2}{*}{Start time}  & \multicolumn{3}{c}{Temporal context [ms]} \\ \cline{3-5}
\multicolumn{2}{c}{} & {250} & {750} & {1500} \\ \hline\hline
\multicolumn{1}{c}{{Blockwise}} & {-}  & {49.4} & {61.6} & 61.2 \\ \hline
\multirow{2}{*}{Fragment}  & Beginning  & \textbf{54.8} & \textbf{63.1} & \textbf{64.8} \\
 & Random  & {43.8} & {49.8} & {55.5} \\ \hline
\end{tabular}
\end{table}
\endgroup

\subsection{Fragment-level versus blockwise identity reassignment}
Tab.~\ref{tab:tab3_start} compares blockwise and fragment-level reassignment using the student model.
We notice that blockwise scores are inferior to fragment-level ones. 
% \RS{This latter approach performs the assignement of the first block of the fragment (starting at the beginning of the fragment).}
This first indicates that the assumption of spatial identity coherence at the fragment level holds on, on the studied temporal contexts. 
{Moreover, this suggests that taking the fragments' beginnings as starting points for embedding extraction may provide advantage for embedding extraction, in which speech overlap is less probable.} 
To verify this, we perform fragment-level reassignment by taking a random start time along the fragments durations instead of always starting from the beginning.
The performance is presented in the last line of Tab.~\ref{tab:tab3_start}. We notice that scores become even inferior to the blockwise reassignment method, which confirms the previously made statement.

Indeed, the degradation of performance indicates a more challenging input for embedding extraction, likely due to an increased presence of speech overlap when taking a random start along the fragment.
Wrong identity reassignment decisions taken on such difficult inputs will, however, have a negative impact on the whole fragment duration. This tends to explain the better scores observed for the blockwise method, in which a wrong decision on one block doesn't impact the decision on the other blocks.

\section{Conclusion}

In this paper, we explored short temporal context speaker embedding extraction for speaker tracking. 
We proposed a Knowledge Distillation (KD) training approach to learn from a pretrained teacher model, given short temporal inputs and using beamforming for more robustness against speech overlap.
We evaluated our distilled speaker embeddings by performing identity reassignment over the output of a neural tracker at the fragment level.
We studied the feasibility of performing blockwise identity reassignment, to go towards a low-latency speaker embedding based tracking system. 
Our results show improved short temporal embedding extraction, with even further improvements on longer temporal contexts when distilling from the teacher's weights.
We further illustrated an increased robustness in the presence of overlapping speakers, although blockwise reassignment scores suggests more room for improvement regarding this matter.
Future work can focus on the design of lighter and more overlap-robust speaker embeddings extractors.
% developing improved reassignment techniques that leverage information from previous blocks, as well as more tailored speaker embeddings extractors, with lighter architectures.
% and capable of directly utilizing multichannel inputs.
% \textcolor{red}{We showed that the fragment-level method benefits from fragments beginnings as a strategic starting points for embedding extraction, with less probability of simultaneous speech.}

% \bibliographystyle{IEEEbib}
\bibliographystyle{ieeetr} 
\bibliography{refs}

\end{document}